\documentclass[12pt]{article}
\usepackage{amssymb} 
\begin{document} 

\vskip2cm

\begin{center}
{\bf ON THE STATUS OF THREE-NEUTRINO  MIXING}

\vspace{0.3cm}
S. M. Bilenky
\footnote {Report at the Meeting NO-VE, Neutrino oscillations at Venice, 24-26
July 2001}\\
\vspace{0.3cm}
{\em Physik-Department, Technische Universit\"at M\"unchen,
James-Franck-Strasse,
D-85748  Garching, Germany \\} 
\vskip0.3cm
{\em Joint Institute for Nuclear Research, Dubna, Russia}\\

\end{center}
\begin{abstract}
All existing solar and atmospheric neutrino oscillation data
are described by the standard two-neutrino formulas with
neutrino mass squared differences that satisfy the hierarchy relation
$\Delta m^{2}_{sol} \ll \Delta m^{2}_{atm}$. We 
discuss this phenomenon from the point of view of three-neutrino mixing.
Possibilities to see effects of three-neutrino mixing and, 
in particular, effects of
CP-violation in the lepton sector depend on the values
 of $|U_{e3}|^2 $ and $\Delta m^{2}_{sol}$.
We present an upper bound of $|U_{e3}|^2 $, which was obtained from a
three-neutrino analysis of CHOOZ data in the case if the solar neutrino
oscillation parameters lie in the LMA allowed region.
\end{abstract}

There exist at present strong evidences in favor of neutrino masses
 and neutrino oscillations that were
obtained in the atmospheric \cite{AS-K} and solar
 \cite{Cl,Kam,GALLEX,GNO,SAGE,S-K,SNO}
neutrino
experiments. 

From the analysis of the existing data it follows that
\begin{enumerate}

\item The data of all solar neutrino experiments can be described by
two-neutrino oscillations, which are
 characterized by two parameters $\Delta m^{2}_{sol}$ and
$\tan^{2}\theta_{sol}$. 
It is assumed also that the fluxes of initial $\nu_{e}$ are given by
the 
Standard Solar Model \cite{BP}.

After the Super-Kamiokande measurement of the day-night asymmetry 
and the spectrum of recoil electrons 
in the process $\nu e \to \nu e $
\cite{S-K}
and the measurement of flux of $\nu_{e}$ on the earth in the SNO experiment
\cite{SNO}
it was established that the large mixing angle MSW allowed regions (LMA, LOW)
are the most preferable ones \cite{OsS-K,Bahcall,Fogli}.
 For the best-fit values of the
oscillation parameters in the LMA region it was found that  \cite{Bahcall}
\begin{equation}
 \Delta m^{2}_{sol} = 4.5 \cdot 10^{-5}\,\rm{eV}^{2};\,~~
 \tan^{2}\theta_{sol} = 4.1 \cdot 10^{-1}\,.
\label{001}
\end{equation}

\item The data of the atmospheric neutrino experiments  are well
described if we assume that $\nu_{\mu}\to\nu_{\tau} $ oscillations take
place . From the analysis of the Super-Kamiokande data 
for the oscillation parameters the following best-fit values were
found \cite{Grew}
\begin{equation}
\Delta m^{2}_{atm}=2.5 \cdot 10^{-3}\rm{eV}^{2};\,~~
\sin^{2}2\theta_{atm} =1 \,.
\label{002}
\end{equation}

\item Thus, from the analysis of the data of the solar and the atmospheric
 neutrino experiments,
it follows that there is hierarchy of
neutrino mass squared differences 
relevant for the oscillations of solar and atmospheric neutrinos
\begin{equation}
\Delta m^{2}_{sol} \ll \Delta m^{2}_{atm}\,.
\label{002a}
\end{equation}

\end{enumerate}

There exist at present indications in favor of
 $\bar \nu_{\mu}\to \bar\nu_{e}$
oscillations that were obtained in the  accelerator LSND experiment
\cite{LSND}. The explanation of these data requires to introduce the 
third neutrino mass squared difference
 $\Delta m^{2}_{LSND}\simeq 1 \rm{eV}^{2}$.
 The LSND data will be checked by the future MiniBooNE
experiment \cite{MiniB}. We will not consider them here.

The minimal number of massive and mixed neutrinos  that
corresponds to three flavor neutrinos  $\nu_{e}$,  $\nu_{\mu}$,
 $\nu_{\tau}$ is equal to three. We will discuss here the results of 
solar and atmospheric
neutrino experiments in the framework of this {\em minimal scheme}. We have
\begin{equation}
\nu_{\alpha L}
=
\sum_{i=1}^{3}
U_{\alpha i}
\,
\nu_{iL}
\label{003}
\,,
\end{equation}
where $\nu_i$ is the field of a neutrino with mass $m_i$,
the index $\alpha$ runs over $e, \mu,\tau$,
and U is the unitary $3 \times3$ PMNS \cite{P,MNS} mixing matrix.

The probability of the transition $\nu_{\alpha}\to\nu_{\alpha '} $
in vacuum is given by the expression (see, for example, \cite{BGG})
\begin{equation}
P(\nu_\alpha\to\nu_{\alpha'})
=
\left|
\sum_{i=1}^{3} U_{{\alpha'} i} \, 
\left( e^{ - i
 \, 
\Delta{m}^2_{i1} \frac {L}{2 E} } \right)
 U_{{\alpha}i}^*\,\right|^2\,.
\label{004}
\end{equation}

Here $ L$ is the distance between a neutrino source and a neutrino detector,
$E$ is the neutrino energy,  and 
$\Delta m^{2}_{i1} = \Delta m^{2}_{i} - \Delta m^{2}_{1} $
(neutrino masses are numerated in such a way that 
$m_1<m_2< m_3$). We will assume that there is hierarchy of neutrino mass
squared differences

\begin{equation}
\Delta m^{2}_{21} \ll \Delta m^{2}_{31}\,.
\label{005}
\end{equation}

Thus, $\Delta m^{2}_{21}$ is relevant
for oscillations of solar neutrinos and $\Delta m^{2}_{31}$
is relevant for oscillations of atmospheric neutrinos.

Let us consider first neutrino oscillations in atmospheric 
and long baseline (LBL) reactor or accelerator
experiments.
In these experiments, $ \frac {L}{E} \lesssim 10^{3}\frac{\rm{m}}{\rm{MeV}}$ 
 and    
\begin{equation}
\Delta{m}^2_{21} \frac {L}{2 E}\ll 1\,.
\label{006}
\end{equation}

Taking into account inequality (\ref{006}) and using
the unitarity relation 
\begin{equation}
\sum_{i=1,2} U_{{\alpha'} i} \, U_{{\alpha}i}^* =\delta_{\alpha\alpha'}-
U_{{\alpha'} 3} \, U_{{\alpha}3}^*
\label{007}
\end{equation}
for the transition probability 
we find 
\begin{equation}
P (\nu_\alpha\to\nu_{\alpha'})
=
\left|
\delta_{\alpha \alpha'}
+
 U_{{\alpha'} 3} \, U_{{\alpha}3}^*
\left( e^{ - i
 \, 
\Delta m^{2}_{31} \frac {L}{2 E} } - 1 \right)
\right|^2\,.
\label{008}
\end{equation}

Thus, if inequality (\ref{006}) is satisfied, transition probabilities in
the atmospheric (LBL) neutrino oscillation experiments are determined by
$\Delta m^{2}_{31}$ and the elements of the mixing matrix $U_{{\alpha}3}$,
which connect flavor neutrinos with the heaviest neutrino
$\nu_{3}$.
For  $\alpha' \neq \alpha$ from (\ref{008}) we have

\begin{equation}
 P(\nu_{\alpha} \to \nu_{\alpha'}) =
\frac {1} {2} {\mathrm A}_{{\alpha'};\alpha} (1 - \cos\, \Delta m^{2}_{31} \frac 
{L} {2E})\,,
\label{009}
\end{equation}

where
\begin{equation} 
{\mathrm A}_{\alpha'; \alpha} = 4 |U_{{\alpha'}3}|^2  
|U_{{\alpha}3}|^2\,.
\label{010}
\end{equation}

For the survival probability from 
(\ref{008}) we obtain 

\begin{equation}
P(\nu_\alpha \to \nu_\alpha) 
= 1 - \frac {1} {2} {\mathrm B}_{\alpha; \alpha}
(1 - \cos\,~ \Delta m^2_{3 1} \frac {L} {2E})\,,
\label{011}
\end{equation}

where 
\begin{equation}
{\mathrm B}_{\alpha; \alpha} =
 4\,~ |U_{{\alpha}3}|^2 (1 -  
|U_{{\alpha}3}|^2)\,.
\label{012}
\end{equation}
Thus, due to the hierarchy (\ref{005}) oscillations of atmospheric  
(LBL) neutrinos are described by the two-neutrino type formulas with
{\it the same} $\Delta m^2_{3 1}${\it for all channels}. The quantities 
${\mathrm A}_{\alpha'; \alpha}$ and ${\mathrm B}_{\alpha ;\alpha}$
are oscillation amplitudes. From the unitarity of the mixing matrix it follows
that they
are connected by the relation
\begin{equation}
{\mathrm B}_{\alpha; \alpha} =
\sum_{\alpha'\neq \alpha}{\mathrm A}_{\alpha'; \alpha}
\label{013}
\end{equation}
and satisfy the inequalities
\begin{equation}
 0 \leq {\mathrm B}_{\alpha \alpha} \leq 1;\,~
0\leq 
 {\mathrm A}_{\alpha' \alpha}\leq 1
\label{014}
\end{equation}
The oscillation amplitudes depend on two parameters, say
$|U_{{\mu}3}|^2$ and $|U_{{\tau}3}|^2 $ (due to unitarity of the mixing matrix
 $|U_{e3}|^2=1 -
|U_{{\mu}3}|^2 -|U_{{\mu}3}|^2$ ).

It is important to stress that
the phase of the matrix elements $U_{{\alpha}3}$ does not enter
 into expression (\ref{009}) for the transition probability.
Thus, if there is hierarchy (\ref{005}), the relation
\begin{equation}
 P(\nu_{\alpha} \to \nu_{\alpha'}) =
 P(\bar \nu_{\alpha} \to \bar \nu_{\alpha'}) 
\label{015}
\end{equation}
is satisfied automatically 
and CP violation in the lepton sector can not be revealed by the investigation
of neutrino oscillations in LBL (atmospheric) neutrino experiments.

{\it The hierarchy of neutrino mass squared differences (\ref{005})
 is the reason
why in the leading approximation the results of the atmospheric
 neutrino oscillation experiments can be described by the standard
two- neutrino formulas}.

Let us consider now solar neutrinos. The probability of solar $\nu_{e}$
to survive in vacuum is given by the expression
\begin{equation}
P^{sol}(\nu_e\to\nu_e)
=
\left|
\sum_{i=1,2}| U_{e i}|^{2} \, 
e^{ - i \, 
\Delta{m}^2_{i1} \frac {L}{2 E} } 
+| U_{e 3}|^{2}e^{ - i \, 
\Delta{m}^2_{31} \frac {L}{2 E} }\right|^{2}\,.
\label{016}
\end{equation}
In order to compare the theory with experimental data 
we must average the survival probability
over neutrino energies and over the region where neutrinos are
produced.
Because of the hierarchy (\ref{005}) 
 the interference between the
 first and the second term in (\ref{016}) disappears
due to averaging  and
 for the averaged survival probability we have
\begin{equation}
P^{sol}(\nu_e\to\nu_e)
=| U_{e 3}|^{4}
+(1 -| U_{e 3}|^{2})^{2}\,~ P^{1,2}(\nu_e\to\nu_e)\,,
\label{017}
\end{equation}

where $ P^{1,2}(\nu_e\to\nu_e)$ is two-neutrino survival probability
that depend on $\Delta{m}^2_{21}$
and the angle $\theta_{12}$ that is determined by the relations

\begin{equation}
\cos^2\theta_{21}
=
\frac{ |U_{e1}|^2 }{\sum_{i=1,2}|U_{ei}|^2 }
\,,
~~~
\sin^2\theta_{21}
= 
\frac{ |U_{e2}|^2 }{\sum_{i=1,2}|U_{ei}|^2 }
\,.
\label{018}
\end{equation}
 
The relation (\ref{017}) is also valid in the case of matter \cite{Schramm}.
In this case the electron density $\rho_{e}$ in the  
effective matter potential must be replaced by
 $(1 -| U_{e 3}|^{2})\rho_{e}$.

No zenith angle dependence of  $\nu_{e}$ events was
observed in the Super-Kamiokande atmospheric neutrino experiment
\cite{AS-K}.
 The data 
of this experiment can be described by $\nu_{\mu}\to \nu_{\tau}$
oscillations with the best-fit value of 
the oscillation amplitude ${\mathrm A}_{\tau; \mu}$ equal to one.
 Thus, the  data of the Super-Kamiokande
experiment are compatible with the small
value of $| U_{e 3}|^{2}$. From the analysis of these data, based on
equations (\ref{009}) and  (\ref{011}),
it was found that $| U_{e 3}|^{2} \leq 0.35 $ \cite{Kj}.

The best upper bound of $| U_{e 3}|^{2}$
can be obtained from the results of the LBL reactor experiments CHOOZ
\cite{CHOOZ} and Palo Verde \cite{PaloV} . In these experiments no 
indications in favor of the disappearance of the reactor $\bar \nu_{e}$ 
were found.
We will consider the results of the CHOOZ experiment. In this
experiment  $\bar \nu_{e}$ from two reactors at the distance about 1 km
from the detector were detected. The data of the experiment were analyzed 
in ref. \cite{CHOOZ}
under the assumption of two-neutrino oscillations and the exclusion
plot in the plane of the parameters $\sin^{2}2 \theta \equiv {\mathrm B}_{e;e}$
and $\Delta m^{2}\equiv \Delta m^{2}_{31} $ was obtained.
From this plot it follows that for a fixed value of $\Delta m^{2}_{31}$
we have
\begin{equation}
{\mathrm B}_{e;e}\leq {\mathrm B}_{e;e}^{0}(\Delta m^{2}_{31})\,,
\label{019}
\end{equation}

From (\ref{010}) and (\ref{019}) it follows that 

\begin{equation}
|U_{e 3}|^{2} \leq  
\frac{1}{2}\,\left(1 - \sqrt{1- B_{e;e}^{0}}\right)
\label{020}
\end{equation}

or
\begin{equation}
|U_{e 3}|^{2} \geq
\frac{1}{2}\,\left(1 + \sqrt{1-B_{e;e}^{0} }\right)\,.
\label{021}
\end{equation}

From the CHOOZ exclusion plot
it can be seen that in the region 
$\Delta m^{2}_{31} \geq 2\cdot10^{-3} \rm{eV}^{2}$
we have $B_{e;e}^{0}\leq 2\cdot 10^{-2}$. If the value of $\Delta m^{2}_{31}$
lies in this region from (\ref{020}) and (\ref{021})
we conclude that the element  $ |U_{e 3}|^{2}$
can be small (inequality (\ref{020})) or large
(inequality
(\ref{021})).

This last possibility is excluded by the
solar neutrino data. In fact,
if $ |U_{e 3}|^{2}$
is close to one, than from the unitarity of the mixing matrix 
it follows that $\sum_{i=1,2} |U_{e i}|^{2}$ is small and the
suppression
of the flux of solar $\nu_{e}$, observed in all solar neutrino
experiments,

cannot be explained by  neutrino oscillations. 
This is also obvious from Eq. (\ref{017}). Thus, from the results 
of the CHOOZ 
and solar neutrino experiments for the upper bound of
 $ |U_{e 3}|^{2}$ we have inequality (\ref{020}). At  
$\Delta m^{2}_{31} = 2.5\cdot10^{-3} \rm{eV}^{2}$ (the
Super-Kamiokande best-fit atmospheric value) we have
\begin{equation}
|U_{e 3}|^{2} \lesssim
4\cdot 10^{-2}\,.
\label{022}
\end{equation}

{\it The smallness of $|U_{e 3}|^{2}$  is the reason 
why in the leading approximation oscillations of solar neutrinos are described
by the standard two-neutrino formula}. 

In the limiting case 
$|U_{e 3}|^{2}= 0$ oscillations of solar and atmospheric (LBL) neutrinos
are decoupled \cite{BG}. In this approximation  
solar neutrino experiments allow to
 obtain information on the values of the parameters $\Delta m^{2}_{21}$ and
$\theta_{12}$,
that characterize oscillations $\nu_{e} \to \nu_{\mu,\tau}$.
The atmospheric (LBL) experiments  allow to obtain information
 on  the values of the parameters 
$\Delta m^{2}_{31}$ and
$\theta_{23}$ that characterize oscillations
 $\nu_{\mu} \to \nu_{\tau}$.

There is, however, no general theoretical reasons for $|U_{e 3}|^{2}$
to be equal to zero. The exact value of the parameter $|U_{e 3}|^{2}$
is of a great interest for further investigation of neutrino mixing.
If  $|U_{e 3}|^{2}$ has a nonzero value and  the 
parameter $\Delta m^{2}_{21}$
is not very small there is a possibility to investigate effects of
three-neutrino mixing and, in particular, fundamental effects of CP-violation
in the lepton sector in the future LBL experiments with neutrinos from Neutrino
factories  and Superbeam facilities 
(see \cite{NuFact,NuBeam}
and references therein).

The upper bound (\ref{022}) was obtained under the assumption
that $\Delta m^{2}_{21}$ is small and does not contribute to the 
$\nu_{e}$ survival probability in LBL experiments. If, however, the LMA is the
solution of the solar neutrino problem the value of the parameter
 $\Delta m^{2}_{21}\,~\frac{L}{2E} $ for $\frac{L}{E}$ values of 
the CHOOZ experiment
can be not so small. 
For example, in ref. \cite{Fogli} the following  ranges of neutrino
oscillation parameters in the LMA region were obtained (99.73 \% CL)
\begin{equation}
 2\cdot 10^{-5}\lesssim \Delta m^{2}_{sol}\lesssim 6 \cdot 10^{-4}\rm{eV}^{2};\,~~
3\cdot 10^{-1}\lesssim \tan^{2}\theta_{sol} \lesssim 1 \,.
\label{022a}
\end{equation}

For the average value of the parameter $\frac{L}{E}$
in the CHOOZ experiment ($\langle\frac{L}{E}\rangle=300$ m/MeV) and
 for the value
 $\Delta m^{2}_{21}=5\cdot10^{-4}\rm{eV}^{2}$, which belong to the
 LMA allowed region (\ref{022a}),
 we have

\begin{equation}
\Delta m^{2}_{21}\,~\frac{L}{2E} \simeq 4\cdot 10^{-1}\,.
\label{023}
\end{equation}

In ref. \cite{BNP} the CHOOZ data were reanalyzed
in the framework of  three-neutrino mixing. New exclusion plots
in the plane of parameters $|U_{e 3}|^{2}$ and $\Delta m^{2}_{31}$  were 
obtained for different values of
 $\Delta m^{2}_{21}$ and $\theta_{12}$, belonging to the LMA region.

From Eq. (\ref{004}) it is not difficult to obtain the following 
expression for three-neutrino
$\nu_{e}$ survival probability in vacuum 

\begin{eqnarray}
\lefteqn{P({\bar \nu_e}\to{\bar \nu_e})} \nonumber\\
&& =\,~~ 1 - 2 \, |U_{e 3}|^2 \left( 1 - |U_{e 3}|^2 \right)
\left( 1 - \cos \frac{ \Delta{m}^2_{31} \, L }{ 2 \, E } \right)
\nonumber \\
&& - \,~~{1\over 2} ( 1 - |U_{e 3}|^2 )^{2}\sin ^{2}2\theta_{sol} \,
\left( 1 - \cos \frac{ \Delta{m}^2_{21} \, L }{ 2 \, E } \right)   \\
\label{024}
& & +\,~~ 2|U_{e 3}|^2 ( 1 - |U_{e 3}|^2 ) \sin^{2}\theta_{sol}\, 
\left(\cos
\left( \frac
{\Delta{m}^2_{31} \, L }{ 2 \, E} - \frac {\Delta{m}^2_{21} \, L }{ 2 \,
E}\right)
-\cos \frac {\Delta{m}^2_{31} \, L }{ 2 \, E} \right)\,. 
\nonumber
\end{eqnarray}

The first term in the right hand side of this expression 
coincides with Eq. (\ref{011}) and Eq. (\ref{012}) (for $\alpha=e$).
The second and third terms are corrections due to
$\Delta{m}^2_{21}$. 

Let us notice  that the values of the second and third terms in
(\ref{024}) are determined by 
$(\Delta m^{2}_{21}\,~\frac{L}{2E})^{2}$ and 
$|U_{e 3}|^{2}\,~ (\Delta m^{2}_{21}\,~\frac{L}{2E})$,
respectively. These terms can give sizable contribution to the 
survival probability 
at large values of $\Delta m^{2}_{21}$.

Expression (\ref{023}) was used in \cite{BNP}
to analyze the CHOOZ data.
It was shown that for $\Delta m^{2}_{21}\geq 2\cdot 10^{-4}\,\rm{eV}^{2}$
the upper bounds on
 $|U_{e 3}|^{2}$
are more stringent than those obtained by the original two-neutrino analysis
of the CHOOZ data in ref. \cite{CHOOZ}.
For the best-fit point of the Super-Kamiokande atmospheric neutrino data
$\Delta m^{2}_{31} = 2.5\cdot 10^{-3}\,\rm{eV}^{2}$ and for
$\sin^{2}\theta_{12}=0.5$
the following bounds were
 obtained

\begin{equation}
|U_{e 3}|^{2}\lesssim 4\cdot 10^{-2}\,,~~\Delta m^{2}_{21} = 1\cdot 10^{-4}
 \rm{eV}^{-4}\,.
\label{025}
\end{equation}

\begin{equation}
|U_{e 3}|^{2}\lesssim 3\cdot 10^{-2}\,,~~\Delta m^{2}_{21} = 4\cdot 10^{-4} \rm{eV}^{-4}\,.
\label{026}
\end{equation}

\begin{equation}
|U_{e 3}|^{2}\lesssim 2\cdot 10^{-2}\,,~~\Delta m^{2}_{21} = 6\cdot 10^{-4} \rm{eV}^{-4}\,.
\label{027}
\end{equation}

These bounds must be compared with the bound (\ref{022})
obtained from the original  two-neutrino analysis 
of the CHOOZ data \cite{CHOOZ}.

With the continuation of the Super-Kamiokande, SNO, GNO and SAGE 
solar neutrino
experiments and with the future BOREXINO experiment \cite{BOREX} a progress in
the establishment of the unique solution of the solar neutrino problem is
expected. The LMA allowed region will be probed in the KamLAND experiment 
\cite{KamL}
scheduled to start in 2001.
After the solution of the solar neutrino problem will be settled and
 oscillation
parameters will be determined with better precision than to-day 
it will be possible 
from the data of the CHOOZ and the Palo Verde experiments to
obtain an exact three-neutrino upper bound of $|U_{e 3}|^{2}$.

In conclusion, we have discussed here the present status of three
neutrino mixing from the point of view of compelling 
 evidences in favor of neutrino
oscillations that were obtained in atmospheric and solar neutrino experiments.
We have demonstrated that due to the hierarchy of neutrino mass squared 
differences  
in the leading approximation neutrino oscillations in
atmospheric and long baseline neutrino experiments  are
described by two-neutrino type formulas.

From the results of the atmospheric neutrino experiments and 
LBL reactor experiments CHOOZ and Palo Verde 
it follows that the element
$|U_{e 3}|^{2}$
that connects  $\nu_{eL}$  with the field of the heaviest neutrino
 $\nu_{3L}$  is small. The smallness of $|U_{e 3}|^{2}$ is the reason why
existing solar neutrino data are described by the standard
two-neutrino formula. 

In the limiting case $|U_{e 3}|^{2}=0$ oscillations of atmospheric (LBL)
neutrinos are $\nu_{\mu} \to\nu_{\tau}$ and oscillations of solar neutrinos
are $\nu_{e} \to \nu_{\mu,\tau}$. In this limiting case $\theta _{13}=0$ and
the investigation of oscillations of 
solar neutrinos allow us to
measure the parameters $\theta _{12}$ and $\Delta m^{2}_{21}$ and
investigation of oscillations of atmospheric (LBL) neutrinos
allow us to measure parameters $\theta _{23}$ and $\Delta m^{2}_{31}$.

There is, however, no general arguments for $|U_{e 3}|^{2}$ to be equal 
to zero. The possibilities to observe effects of three-neutrino mixing  
in atmospheric (LBL) neutrino experiments depend on the value of $|U_{e
3}|^{2}$ and on the value of  $\Delta m^{2}_{21}$. 
For the detailed discussion of the sensitivity of the future neutrino
oscillation experiments to
the quantity 
 $|U_{e 3}|^{2}$
and to the CP-phase $\delta$ see report \cite{ARub} at this meeting.

I acknowledge the Alexander von Humboldt Foundation for support.


\begin{thebibliography}{99}


\bibitem{AS-K} Super-Kamiokande Collaboration, S.~Fukuda {\it et al.,}
Phys. Rev. Lett. {\bf 81}, 1562 (1998),
 S.~Fukuda {\it et al.,} Phys. Rev. Lett. {\bf 82}, 2644 (1999),
S.~Fukuda {\it et al.,} Phys. Rev. Lett.{\bf 85}, 3999-4003 (2000).



\bibitem{Cl}B. T. Cleveland {\it et al.}, Astrophys. J. {\bf
496}, 505 (1998).

\bibitem{Kam} Kamiokande Collaboration, Y. Fukuda {\it et al.},
  Phys. Rev. Lett. {\bf 77}, 1683 (1996).

\bibitem{GALLEX} GALLEX Collaboration, W. Hampel {\it et al.}, 
 Phys. Lett. B {\bf 447}, 127 (1999).

\bibitem{GNO} GNO Collaboration,
M. Altmann {\it et al.},  
Phys. Lett. B {\bf 490}, 16 (2000).

\bibitem{SAGE} SAGE Collaboration, J. N. Abdurashitov {\it et al.},
 Phys. Rev. C {\bf 60}, 055801 (1999).


\bibitem{S-K} Super-Kamiokande Collaboration, S.~Fukuda {\it et al.},
Phys. Rev. Lett. {\bf 86}, 5651 (2001).

\bibitem{SNO}SNO collaboration Q.R. Ahmad {\it et al.}, 
Phys. Rev. Lett. {\bf 87}, 071301 (2001).


\bibitem{BP}J. N. Bahcall, M. H. Pinsonneault, and S. Basu,
Astrophys. J. {\bf 555}, 990 (2001).

\bibitem{OsS-K}  S.~Fukuda {\it et al.},
Phys. Rev. Lett. {\bf 86}, 5656 (2001).

\bibitem{Bahcall} J. N. Bahcall, M. C. Gonzalez-Carcia and C.Pena-Garay,
hep-ph/0106258.

\bibitem{Fogli} G. L. Fogli, E. Lisi,  D. Montanino, and  A. Palazzo,
 hep-ph/0106247.


\bibitem{Grew} Super-Kamiokande Collaboration, C. McGrew,
Proceeding of
the IXth International Workshop on "Neutrino Telescopes"( Venice, Italy, 
March 6-9, 2001).
                         
                                             
\bibitem{BOR} BOREXINO Collaboration,
               report by G.\ Ranucci, Proceedings of the
19th International Conference on Neutrino
                Physics and Astrophysics,
{\em Neutrino~2000\/}\,(Sudbury, Canada, June 16-21, 2000).


\bibitem{LSND} LSND Collaboration,
               report by G.\ Mills, Proceedings of the
19th International Conference on Neutrino
                Physics and Astrophysics,
{\em Neutrino~2000\/}\,(Sudbury, Canada, June 16-21, 2000).

\bibitem{MiniB} MiniBooNE Collaboration, report by A.\ Bazarko,
Proceedings of the
19th International Conference on Neutrino Physics and Astrophysics,
{\em Neutrino~2000\/}\,(Sudbury, Canada, June 16-21, 2000).
\bibitem{P}
B.~Pontecorvo,
J. Exptl. Theoret. Phys. \textbf{34}, 247 (1958)
[Sov. Phys. JETP \textbf{7}, 172 (1958)];
B.~Pontecorvo,
Zh. Eksp. Teor. Fiz. \textbf{53}, 1717 (1967)
[Sov. Phys. JETP \textbf{26}, 984 (1968)].

\bibitem{MNS}
Z.~Maki, M.~Nakagawa, and S.~Sakata,
Prog. Theor. Phys. \textbf{28}, 870 (1962).

\bibitem{BGG} S.M.\, Bilenky, C.\, Giunti and
W.\,Grimus. Prog. Part. Nucl. Phys. {\bf 43}, 1 (1999), hep-ph/9812360.

\bibitem{Schramm} S.T. Petcov, 
Phys. Lett. B{\bf 214} (1988) 259;
 X.~Shi and D.N.~ Schramm,
Phys. Lett. B \textbf{283}, 305 (1992).

\bibitem{Kj} T.\, Kajita, report at 
Neutrino Oscillation Workshop
{\em NOW2000}, Otranto, Italy, Sept. 9-16, 2000;
G.L. Fogli et al., hep-ph/0104221.

 \bibitem{CHOOZ} CHOOZ Collaboration, M.\, Apollonio \textit{et al},
Phys. Lett. B{\bf 338}, 383 (1998);
 M.\, Apollonio \textit{et al.},
Phys. Lett. B{\bf 466} (1999) 415.

\bibitem{PaloV} F.\, Boehm, J. \textit{ et al.},
Phys.\ Rev.\ Lett.\  {\bf 84}, 3764 (2000); 
Phys. Rev. D{\bf 62} (2000) 072002.


\bibitem{BG} S.M.Bilenky and C.Giunti, Phys. Lett. B{\bf 444} (1998) 379.

\bibitem{NuFact} 
C.~Albright {\em et al.} "Physics at a Neutrino Factory" hep-ex/0008064;
A.~Blondel {\em et al.} "The Neutrino factory: beam and experiments",

\bibitem{NuBeam} V.\, Barger \textit{et al.}, hep-ph/0103052;
J.\, Gomes Cadenas,
Proceedings of the 9th  International Workshop "Neutrino Telescopes"
 (Venice, March 6-9, 2001).
\bibitem{BNP} S.M.Bilenky, D.Nicolo and S.T.Petcov in preparation.

\bibitem{BOREX} BOREXINO Collaboration,
 G.\ Ranucci, Proceedings of the
19th International Conference on Neutrino
                Physics and Astrophysics,
{\em Neutrino~2000\/}\,(Sudbury, Canada, June 16-21, 2000).

\bibitem{KamL}  KamLAND Collaboration, A. Piepke,
Proceedings of the
19th International Conference on Neutrino
                Physics and Astrophysics,
{\em Neutrino~2000\/} (Sudbury, Canada, 2000).

\bibitem{ARub} A.Rubbia, Proceedings of the International Workshop NO-VE
"Neutrino Oscillations in Venice" (Venice, July 24-26, 20001).






\end{thebibliography}
\end{document}